\documentclass[12pt]{article}
\usepackage{a4wide,epsfig,psfrag,amsmath,amssymb,cite,scalefnt}
\usepackage{color}

\newcommand{\gsim}{\;\rlap{\lower 3.5 pt \hbox{$\mathchar \sim$}} \raise 1pt
 \hbox {$>$}\;}
\newcommand{\lsim}{\;\rlap{\lower 3.5 pt \hbox{$\mathchar \sim$}} \raise 1pt
 \hbox {$<$}\;}

\newcommand{\mgluino}{m_{\tilde{g}}}
\newcommand{\mstopone}{m_{\tilde{t}_1}}
\newcommand{\mstoptwo}{m_{\tilde{t}_2}}
\newcommand{\msquark}{m_{\tilde{q}}}
\newcommand{\msusy}{m_{\rm SUSY}}

\allowdisplaybreaks

\begin{document}

\title{\vskip-3cm{\baselineskip14pt
    \begin{flushleft}
      \normalsize SFB/CPP-12-30\\
      \normalsize TTP12-22\\
      \normalsize LPN12-068 
  \end{flushleft}}
  \vskip1.5cm
  Decoupling constant for $\alpha_s$ and the 
  effective gluon-Higgs coupling to three loops
  in supersymmetric QCD
}

\author{
  Alexander Kurz,
  Matthias Steinhauser,
  Nikolai Zerf
  \\[1em]
  {\small\it Institut f{\"u}r Theoretische Teilchenphysik}\\
  {\small\it Karlsruhe Institute of Technology (KIT)}\\
  {\small\it 76128 Karlsruhe, Germany}
}

\date{}

\maketitle

\thispagestyle{empty}

\begin{abstract}
  We compute the three-loop QCD corrections to the decoupling constant for
  $\alpha_s$ which relates the Minimal Supersymmetric Standard Model to
  Quantum Chromodynamics with five or six active flavours. The new results can
  be used to study the stability of $\alpha_s$ evaluated at a high scale from
  the knowledge of its value at $M_Z$. We furthermore derive a low-energy
  theorem which allows the calculation of the coefficient function of the
  effective Higgs boson-gluon operator from the decoupling constant. 
  This constitutes the first independent check of the matching coefficient to
  three loops.
  \medskip

  \noindent
  PACS numbers: 11.30.Pb 12.38.Bx

\end{abstract}

\thispagestyle{empty}


\newpage


\section{Introduction}

The decoupling of particles with masses much heavier than the considered
energy scale has a long history~\cite{Weinberg:1980wa}. It is tightly
connected to the construction of an effective theory containing only the light
active degrees of freedom in the dynamical part of the Lagrange
density. Within the framework of QCD decoupling constants for the strong
coupling $\alpha_s$ are known at
two-~\cite{Bernreuther:1981sg,Larin:1994va,Chetyrkin:1997un},
three~\cite{Chetyrkin:1997un} and even four-loop
order~\cite{Schroder:2005hy,Chetyrkin:2005ia}. Recently also the expression
for the simultaneous decoupling of two heavy quarks has been computed at the
three-loop level~\cite{Grozin:2011nk}.

Decoupling relations are also important in the context of supersymmetry where
the Standard Model constitutes the effective theory. Two-loop corrections for
a degenerate supersymmetric mass spectrum are known from
Ref.~\cite{Harlander:2005wm,Harlander:2007wh} and the general result can be found in
Ref.~\cite{Bauer:2008bj}. In this paper we compute
the three-loop corrections 
for several different assumptions on the masses of the MSSM.

There is an interesting connection between the decoupling constants and
the effective coupling of a CP neutral Higgs boson to gluons which is
defined via the Lagrange density (the superscript $0$ marks bare
quantities) 
\begin{eqnarray}
  {\cal L}_{Y,\rm eff} &=& -\frac{\phi^0}{v^0} C_1^0 {\cal O}_1^0 
  + {\cal L}_{QCD}^{(5)}
  \,,
  \label{eq::leff}
\end{eqnarray}
with
\begin{eqnarray}
  {\cal O}_1^0 &=& \frac{1}{4} G_{\mu\nu}^0 G^{0,\mu\nu}\,,
\end{eqnarray}
where $\phi$ is the Higgs field $v$ is the vacuum expectation value and
$G_{\mu\nu}$ the field strength tensor in QCD. ${\cal L}_{QCD}^{(5)}$ is the
QCD Lagrange density with five active flavours.
The first term in Eq.~(\ref{eq::leff}) describes the coupling of the Higgs
boson to two, three and four gluons.

In Ref.~\cite{Chetyrkin:1997un} an all-order low-energy theorem (LET)
has been derived which connects $C_1$ to the derivative of the decoupling
constant for $\alpha_s$ with respect to the top quark
mass.\footnote{Discussions about the LET applied at one and two loops can,
  e.g., be found in Refs.~\cite{Ellis:1975ap,Shifman:1979eb,Kniehl:1995tn}.}
As far as supersymmetry is concerned a next-to-leading order (NLO) version of
the LET has been derived in Ref.~\cite{Degrassi:2008zj} (see also
Ref.~\cite{Mihaila:2010mp}). In this way the NLO supersymmetric QCD (SQCD)
corrections to $C_1$ obtained in Ref.~\cite{Harlander:2004tp} could be
confirmed.  We re-derive the LET, apply it at three loops and thus obtain the
coefficient function $C_1$ which is needed for NNLO prediction of Higgs boson
production and decay within the MSSM.  With our calculation we confirm the
result for $C_1$ obtained in Ref.~\cite{Pak:2010cu,PSZ_2012} by an explicit
calculation of the vertex diagrams.

The outline of this paper is as follows: In the next Section we describe the
calculation of the decoupling constant for $\alpha_s$ to three loops and
discuss the numerical influence in the computation of 
$\alpha_s(M_{\rm GUT})$. Afterwards we derive in Section~\ref{sec::let} an
all-order low-energy-theorem which we use to compute $C_1$ to NNLO accuracy.
We summarize and conclude the paper in
Section~\ref{sec::concl}. In the Appendix we present a compact expression of
the exact two-loop result for the decoupling coefficient.


\section{\label{sec::dec}Decoupling of heavy supersymmetric particles}

In order to compute the decoupling effects of heavy particles from the running
of $\alpha_s$ one can use the well-established formalism derived in
Ref.~\cite{Chetyrkin:1997un}. It has been applied to supersymmetry in
Refs.~\cite{Harlander:2005wm,Harlander:2007wh,Bauer:2008bj} where two-loop
corrections have been computed.

The starting point is the relation between the strong coupling in the full
theory, which is in our case the MSSM, respectively, SQCD, and the effective
theory, QCD
\begin{eqnarray}
  \alpha_s^{\rm (QCD)}(\mu) = \zeta_{\alpha_s}(\mu) \alpha_s^{\rm (SQCD)}(\mu)
  \,.
  \label{eq::zetaas}
\end{eqnarray}
At that point some comments are in order:
\begin{itemize}
\item $\alpha_s^{\rm (QCD)}(\mu)$ is defined in the five or six flavour
  theory, depending on whether the top quark is integrated out together with
  the supersymmetric particles or not.
\item $\alpha_s^{\rm (QCD)}(\mu)$ is defined in the $\overline{\rm MS}$ scheme
  based on Dimensional Regularization (DREG).  $\alpha_s^{\rm (SQCD)}(\mu)$ is
  defined in the $\overline{\rm DR}$ scheme since the supersymmetric theory is
  regularized using Dimensional Reduction (DRED).  DRED is implemented with
  $\varepsilon$ scalars, where the details can be found in
  Refs.~\cite{Harlander:2009mn,PSZ_2012}.
\item $\zeta_{\alpha_s}(\mu)$ as introduced in Eq.~(\ref{eq::zetaas}) has two
  tasks: (i) it has to decouple the heavy particles not present in the effective
  theory, and (ii) $\zeta_{\alpha_s}(\mu)$ has to ensure the change of 
  regularization from DRED to DREG.
  In principle the two tasks can be performed in two steps as it has been
  proposed in
  Refs.~\cite{Harlander:2005wm,Harlander:2007wh,Bauer:2008bj}. However, it is
  more convenient
  to choose the same renormalization scale for the decoupling and the change
  of scheme. Calculations along these lines have also been performed in
  Ref.~\cite{Bednyakov:2007vm,Pak:2010cu}. 
\item In principle each vertex containing $\alpha_s$ can be used in order to
  compute $\zeta_{\alpha_s}$.  It is, however, convenient to use the
  gluon-ghost vertex in order to compute the decoupling constant
  via~\cite{Chetyrkin:1997un}
  \begin{eqnarray}
    \zeta_{\alpha_s}^0 &=&
    \left(\frac{\tilde\zeta_1^0}{\tilde\zeta_3^0\sqrt{\zeta_3^0}}\right)^2
    \,,
    \label{eq::zeta_bare}
  \end{eqnarray}
  where the superscript ``0'' marks bare quantities.
  $\tilde\zeta_1$, $\tilde\zeta_3$ and $\zeta_3$ are the decoupling
  constants of the gluon-ghost vertex, ghost and gluon propagator,
  respectively. They are obtained from the hard part of the corresponding
  Green's function 
    (see Fig.~\ref{fig::diags} for sample Feynman diagrams up
  to three loops). The corresponding formulae can be found in
  Ref.~\cite{Chetyrkin:1997un} where a derivation has been performed in the 
  framework of QCD. It can be taken over to SQCD without modifications.
  The renormalized decoupling constant is obtained from
  \begin{eqnarray}
    \zeta_{\alpha_s} &=& 
    \frac{Z_{\alpha_s}}{Z_{\alpha_s^\prime}}\zeta_{\alpha_s}^0
    \,,
  \end{eqnarray}
  where $Z_{\alpha_s}$ and $Z_{\alpha_s^\prime}$ are the renormalization
  constants for $\alpha_s$ in the full and effective theory, respectively.

\begin{figure}[t]
  \centering
  \includegraphics[width=\linewidth]{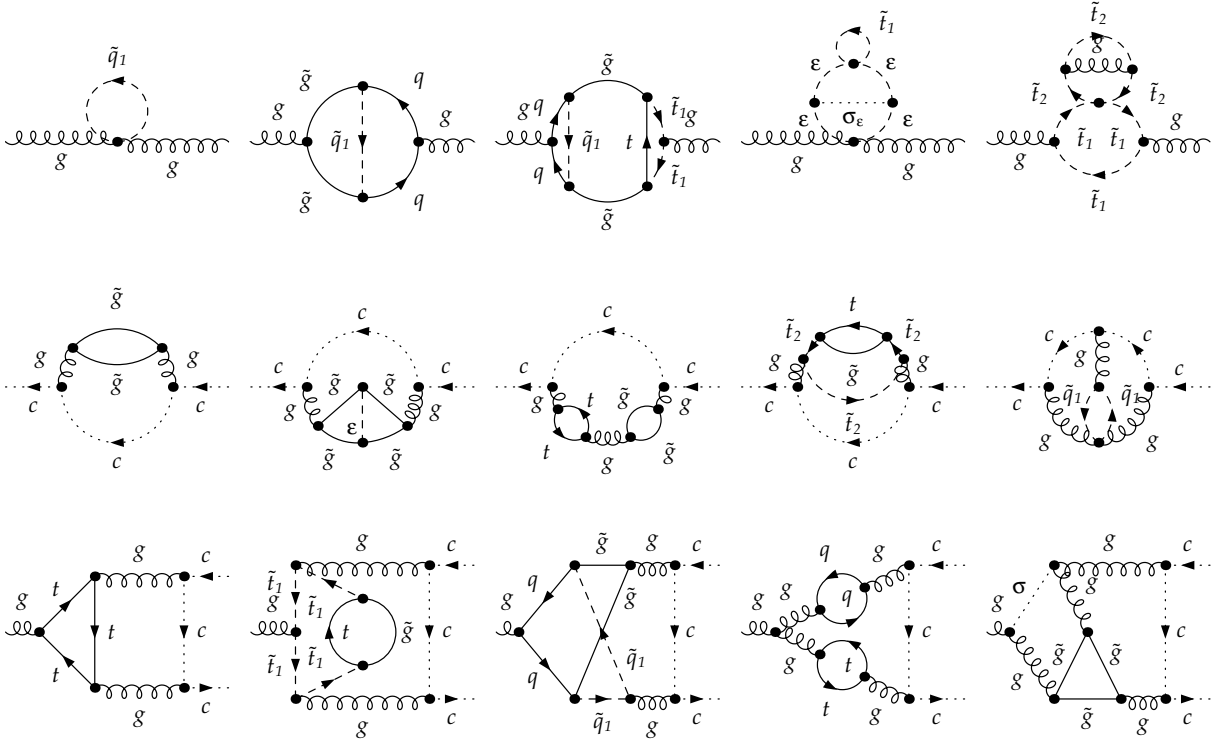}
  \caption[]{\label{fig::diags}Sample diagrams contributing to 
    $\zeta_3$ (top row), $\tilde\zeta_3$ (middle row)
    and $\tilde\zeta_1$ (bottom row)
    up to three loops. 
    The symbols $t$, $\tilde{t}_i$, $q$, $\tilde{q}_i$, 
    $g$, $\tilde{g}$, $c$ and
    $\varepsilon$ denote top quarks, top squarks, light quarks and the
    corresponding squarks, gluons, gluinos, ghosts
    and $\varepsilon$ scalars, respectively.
    $\sigma_\varepsilon$ and $\sigma$ are auxiliary particles used for the implementation
    of the four-$\varepsilon$ and four-gluon vertices, respectively.
    }
\end{figure}

\item All occurring parameters are renormalized in the $\overline{\rm DR}$
  scheme, except the $\varepsilon$ scalar mass which is renormalized on-shell
  with the condition $M_\varepsilon=0$. The corresponding counterterms can,
  e.g., be found in Ref.~\cite{Hermann:2011ha}.
\end{itemize}

Assuming a strong hierarchy among the quarks one encounters in the case of QCD
vacuum diagrams which contain only one mass scale.  The occurring integrals
can even be computed up to four-loop
order~\cite{Schroder:2005hy,Chetyrkin:2005ia}. Two scales appear if two quarks
are integrated out simultaneously. This has been done in
Ref.~\cite{Grozin:2011nk} to three-loop accuracy.

In the case of supersymmetry significantly more mass scales have to be
considered. In our approach we have the gluino and top squark masses
($\mgluino$, $\mstopone$, $\mstoptwo$) and a generic squark mass $\msquark$
which we take as the average of the up, down, strange, charm and bottom
squarks.  In addition there is the $\varepsilon$ scalar ($M_\varepsilon$) and
the top quark ($m_t$) mass. The latter only appears if we match to
five-flavour QCD since $m_t=0$ is chosen for the matching to six-flavour
  QCD.  Up to two loops $\zeta_{\alpha_s}$ can nevertheless be computed
exactly~\cite{Bauer:2008bj} taking into account the dependence on all mass
parameters. The analytical result can be found in the Appendix. At three-loop
order, however, approximations have to be adopted in order to be able to
compute the integrals. Motivated by scenarios which are currently discussed in
the literature we have chosen
\begin{eqnarray}
  ({\rm h1}) && \msquark \approx \mstopone \approx \mstoptwo \approx \mgluino \gg m_t
  \,,\nonumber\\
  ({\rm h2}) && \msquark \approx \mstoptwo \approx \mgluino \gg \mstopone \gg m_t
  \,,\nonumber\\
  ({\rm h3}) && \msquark \approx \mstoptwo \approx \mgluino \gg \mstopone \approx m_t
  \,,
  \label{eq::hierarchies}
\end{eqnarray}
where in the case of ``$\gg$'' an asymptotic expansion in the corresponding
hierarchy is performed. In the case of ``$\approx$'' a naive Taylor expansion
in the difference of the particle masses is sufficient. For all hierarchies we
assume that $M_\varepsilon$ is not zero but much smaller than all other
masses. In this way we ensure that the $\varepsilon$ scalar is integrated out
and not present in the effective theory. Thus, in the latter dimensional
regularization can be used.  In what follows the heavy mass scales for each
hierarchy are also denoted by $\msusy$ in case they are identified.

Whereas at one- and two-loop order only 12 and 362 Feynman diagrams have to be
considered there are more than 20\,000 at three-loop order. It goes without
saying that it is thus necessary to automate the calculation as much as
possible. We rely on a chain of programs which work hand-in-hand in order to
minimize the error-prone manual interaction: All Feynman diagrams are
generated with {\tt QGRAF}~\cite{Nogueira:1991ex} and afterwards transformed
to {\tt FORM}~\cite{Vermaseren:2000nd} notation with the help of {\tt
  q2e}. The rules of asymptotic expansion (see, e.g.,
Ref.~\cite{Smirnov:2002pj}) are applied on a diagrammatic level using {\tt
  exp}~\cite{Harlander:1997zb,Seidensticker:1999bb} and finally we evaluate
the resulting vacuum integrals which (after asymptotic expansion) only contain
a single scale with the help of the package {\tt
  MATAD}~\cite{Steinhauser:2000ry}.  The automated setup allows us to perform
the calculation for general gauge parameter $\xi$. Whereas $\tilde\zeta_1$,
$\tilde\zeta_3$ and $\zeta_3$ individually depend on $\xi$ it drops out in the
combination for $\zeta_{\alpha_s}$ which serves as a welcome check for our
calculation.  A further check is provided by the overlap of the numerical
results of the three hierarchies defined in Eq.~(\ref{eq::hierarchies}) as we
will discuss below.

At three-loop order terms up to ${\cal O}(1/\msusy^{10})$ have been computed
for (h1) and (h3) and up to ${\cal O}(1/\mstopone^{6})$ and ${\cal
  O}(1/\msusy^{6})$ for (h2). For each mass difference at least four expansion
terms (i.e. terms including $(m_i^2-m_j^2)^3$) could be evaluated. It is
either possible to expand in the linear or the quadratic mass
difference. Formally both choices are equivalent, however, in practice it
turns out that depending on the actual numerical values of the parameters one
can be significantly better behaved than the other. Similarly there is a
freedom to choose a mass parameter, $m_R$, around which the expansion is
performed. $m_R$ should be of the order of the involved masses. Note that
for (h1) and (h2) only one reference mass $m_R$ is required whereas for (h3)
one needs two as can be seen from Eq.~(\ref{eq::hierarchies}).  Again there
may be significant numerical differences and thus we adopt the following
choices when evaluating the three-loop corrections to the decoupling
coefficient
\begin{eqnarray}
  ({\rm h1}) && m_R = \mstopone\,, m_R = \mstoptwo\,, 
  m_R = \mgluino\,, m_R = \msquark\,,
  m_R = \frac{\mstopone+\mstoptwo+10\msquark+\mgluino}{13}\,,
  \nonumber\\
  ({\rm h2}) && m_R = \mstoptwo\,, 
  m_R = \mgluino\,, m_R = \msquark\,,
  m_R = \frac{\mstoptwo+10\msquark+\mgluino}{12}\,,
  \nonumber\\
  ({\rm h3}) && m_{R_1} = \mstoptwo\,, 
  m_{R_1} = \mgluino\,, m_{R_1} = \msquark\,,
  m_{R_1} = \frac{\mstoptwo+10\msquark+\mgluino}{12}\,,
  \nonumber\\&&\mbox{}
  m_{R_2} = m_t\,, 
  m_{R_2} = \mstopone\,,
  m_{R_2} = \frac{m_t + \mstopone}{2}\,.
  \label{eq::mR}
\end{eqnarray}

In the following it is convenient to consider the perturbative expansion of
$\zeta_{\alpha_s}$ which we define as
\begin{eqnarray}
  \zeta_{\alpha_s}(\mu) &=& 1 
  + \frac{\alpha_s^{\rm (SQCD)}}{\pi} \zeta_{\alpha_s}^{(1)}
  + \left(\frac{\alpha_s^{\rm (SQCD)}}{\pi}\right)^2 \zeta_{\alpha_s}^{(2)}
  + \left(\frac{\alpha_s^{\rm (SQCD)}}{\pi}\right)^3 \zeta_{\alpha_s}^{(3)}
  + \ldots\,,
  \label{eq::zetaasexp}
\end{eqnarray}
where the $\mu$ dependence of $\alpha_s^{\rm (SQCD)}$ and
$\zeta_{\alpha_s}^{(i)}$ is suppressed on the right-hand side.

The general results are quite lengthy and will not be presented in this
paper. However, in order to get an impression of the results we present
$\zeta_{\alpha_s}$ for the hierarchy (h1) with a degenerate supersymmetric
mass spectrum which reads
\begin{eqnarray}
  \zeta_{\alpha_s}^{(1)} &=& -\frac{1}{4} - l_S - \frac{l_t}{6}
  \,,\nonumber\\
  \zeta_{\alpha_s}^{(2)} &=& 
  \frac{307}{288} 
  + \left(-\frac{77}{72} + \frac{7}{3}l_x\right) l_t
  + \frac{49}{36}l_t^2
  - \frac{25}{36}l_x
  + l_x^2
  + x_{tS} \left(\frac{1}{432} + \frac{1}{9}l_t + \frac{13}{72}l_x\right)
  \nonumber\\ &&\mbox{}
  + x_{tS}^2 \left(-\frac{1597}{21600} + \frac{61}{720} l_x\right)
  + \ldots
  \,,\nonumber\\
  \zeta_{\alpha_s}^{(3)} &=& 
  \frac{162443}{62208}
  - \frac{8509}{3456}\zeta(3) 
  + \left(-\frac{27013}{5184} + \frac{2581}{432}l_x - \frac{7}{2}l_x^2\right) l_t
  + \left(\frac{6361}{1728} - \frac{49}{12}l_x \right) l_t^2
  \nonumber\\ &&\mbox{}
  - \frac{343}{216}l_t^3
  - \frac{21583}{5184}l_x
  + \frac{641}{288} l_x^2
  - l_x^3
  + x_{tS} \left(-\frac{90481643}{3888000} 
    + \frac{47429}{2304}\zeta(3)
    \right.\nonumber\\ && \left.\mbox{}
    + \left(\frac{12163}{21600} - \frac{122}{135}l_x \right) l_t
    - \frac{79}{216} l_t^2
    + \frac{51353}{86400} l_x
    - \frac{69}{128}l_x^2
  \right)
  + x_{tS}^2
  \left(\frac{1542497350769}{64012032000} 
    \right.\nonumber\\ && \left.\mbox{}
    - \frac{2330095}{110592}\zeta(3)
    + \left(\frac{585083}{12700800} - \frac{26807}{60480}l_x\right) l_t
    - \frac{2}{27}l_t^2
    + \frac{3208403}{3386880}l_x
    - \frac{104479}{181440} l_x^2
    \right)
    \nonumber\\ &&\mbox{}
    +\ldots
  \,,
  \label{eq::zeta_deg}
\end{eqnarray} 
where $x_{tS}=m_t^2/\msusy^2$, 
$l_t=\ln(\mu^2/m_t^2)$, $l_S=\ln(\mu^2/\msusy^2)$ and $l_x=\ln(x_{tS})$.
The ellipses denote terms of order $x_{tS}^3$.  
The corresponding results where the matching is performed to six-flavour QCD,
i.e. where the top quark is not integrated out and thus treated as massless in
the loop integrals, reads
\begin{eqnarray}
  \zeta_{\alpha_s}^{(1)} &=& -\frac{1}{4} - l_S
  \,,\nonumber\\
  \zeta_{\alpha_s}^{(2)} &=& \frac{77}{96} - \frac{7}{12} l_S + l_S^2
  \,,\nonumber\\
  \zeta_{\alpha_s}^{(3)} &=& -\frac{11203}{4608} - \frac{1495}{576} l_S +
  \frac{541}{288} l_S^2 - l_S^3 + \frac{4657}{9216} \zeta (3) 
  \,.
  \label{eq::zeta_deg2}
\end{eqnarray} 
All analytical expressions corresponding to the hierarchies of
Eq.~(\ref{eq::hierarchies}) can be found in the file {\tt decsusy3l.m} 
obtained from~\cite{progdata}.

\begin{figure}[t]
  \centering
  \includegraphics[width=.8\linewidth]{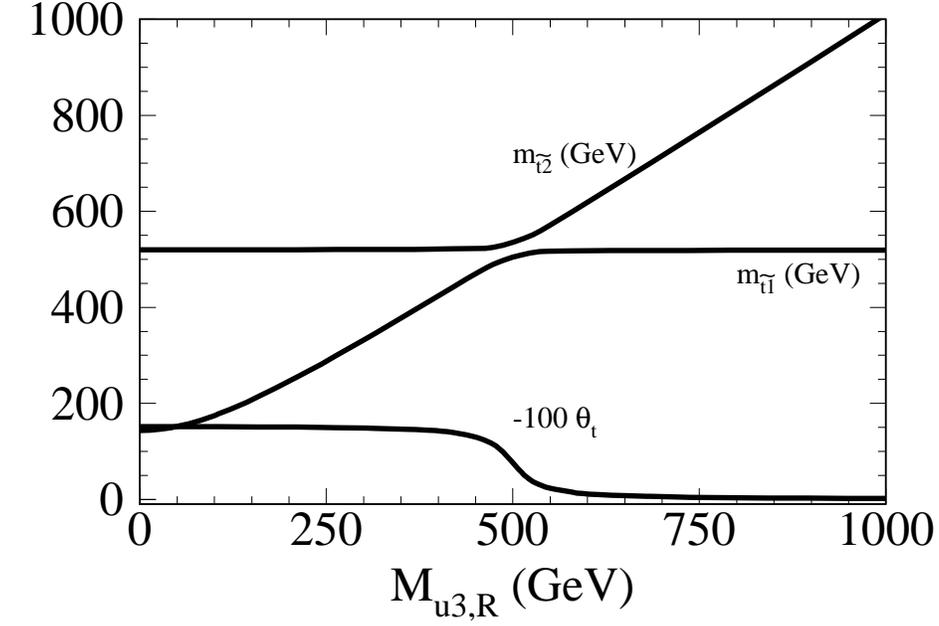} 
  \\ 
  (a)
  \\ [-1em]
  \includegraphics[width=.8\linewidth]{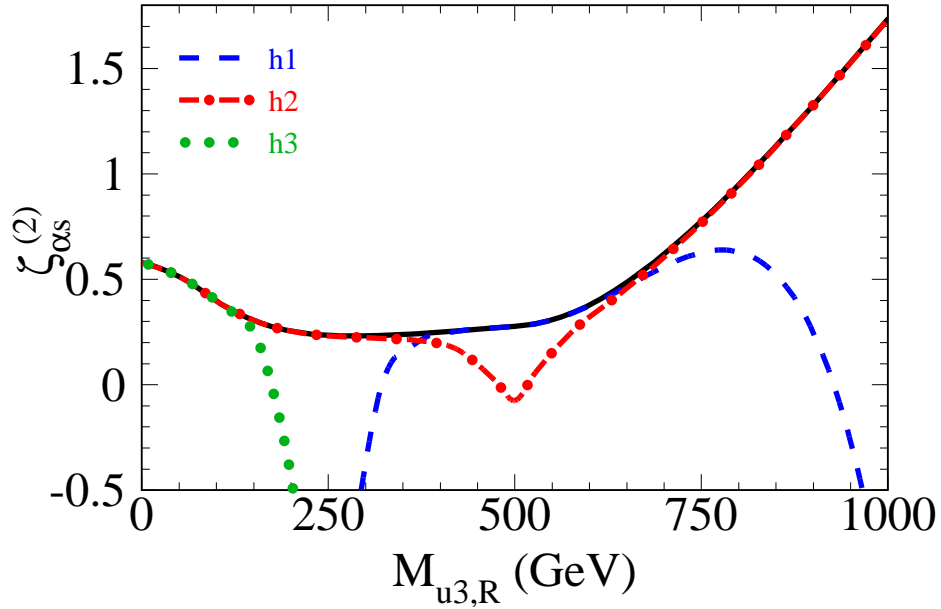}
  \\ 
  (b)
  \caption[]{\label{fig::zeta2l}(a) $\mstopone$, $\mstoptwo$ and
    $\theta_{t}$ obtained from the diagonalization of the top
      squark mass matrix as a function of the soft SUSY breaking parameter
    $M_{\tilde{u}_3,R}$. (b) $\zeta_{\alpha_s}^{(2)}$ as a function of
    $M_{\tilde{u}_3,R}$ using the parameters of Eq.~(\ref{eq::parameters}).
    The exact result is shown as solid black line.}
\end{figure}

Let us in the following test our approximation at two loops by comparing to
the exact result. For this purpose we adopt the following values for the input
parameters
\begin{align}
  &&m_t=150~\mbox{GeV}\,,
  &&A_t=100~\mbox{GeV}\,,
  &&M_{\tilde Q_3}=500~\mbox{GeV}\,,
  &&\mu_{\rm SUSY}=100~\mbox{GeV}\,,
  \nonumber\\
  &&\tan\beta = 10\,,
  &&M_Z = 91.2~\mbox{GeV}\,,
  &&\sin^2\theta_W=0.2233\,,
  \label{eq::parameters}
\end{align}
where $A_t$ is the trilinear coupling, $\mu_{\rm SUSY}$ is the Higgs-Higgsino
bilinear coupling from the super potential, $\tan\beta$ is the ratio of the
vacuum expectation values of the two Higgs doublets, $M_Z$ is the $Z$ boson
mass, $\theta_W$ the weak mixing angle and $M_{\tilde{Q}_3}$, a soft SUSY
breaking parameter for the squark doublet of the third family. Furthermore we
set the renormalization scale to $\mu=500$~GeV.  These parameters can be used
to compute $\mstopone$, $\mstoptwo$ and $\theta_{t}$ as a function of
the singlet soft SUSY breaking parameter of the top squark,
$M_{\tilde{u}_3,R}$ (see, e.g., Ref.~\cite{Martin:1997ns}) by
diagonalizing the corresponding mass matrix. The result is
shown in Fig.~\ref{fig::zeta2l}(a). Furthermore we choose for simplicity
$\mstoptwo=\msquark=\mgluino$.  This allows us to consider in
Fig.~\ref{fig::zeta2l}(b) both the exact result for $\zeta_{\alpha_s}^{(2)}$
(solid line) and the approximations (dahed lines) based on the hierarchies
(h1), (h2) and (h3). The latter are obtained from the (naive)
averages over the
various representations, i.e., the different choices of $m_R$ according to
Eq.~(\ref{eq::mR}).  One observes that in the whole range of
$M_{\tilde{u}_3,R}$ at least one of the hierarchies approximates the exact to
a high degree, which provides the motivation to proceed in a similar way at
three loops.

\begin{figure}[t]
  \centering
  \includegraphics[width=1.\linewidth]{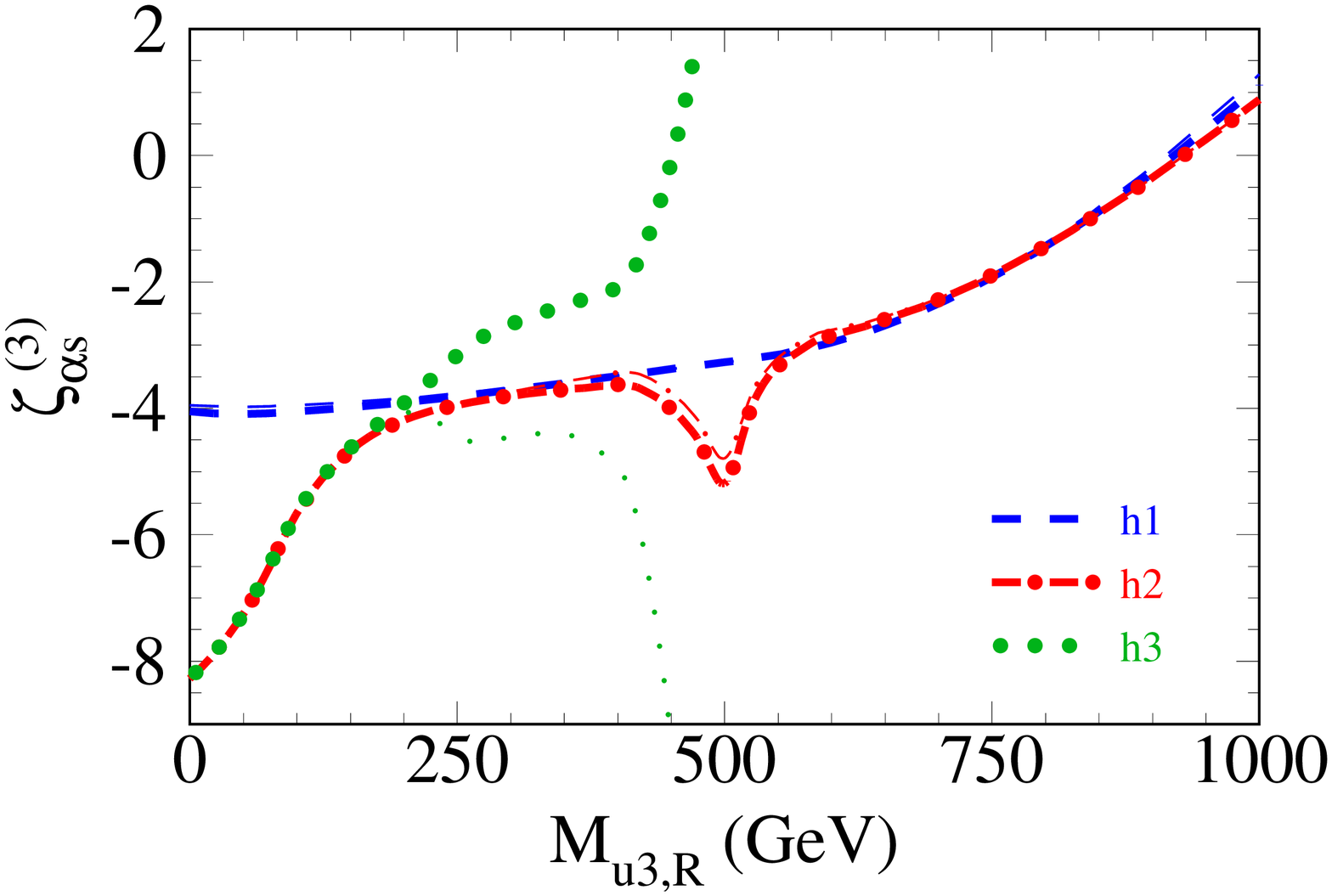}
  \caption[]{\label{fig::zeta3l}$\zeta_{\alpha_s}^{(3)}$ as a function of
    $M_{\tilde{u}_3,R}$ using the parameters of Eq.~(\ref{eq::parameters}).
    Thick lines include all available terms whereas for the thin lines the
    highest terms are cancelled.}
\end{figure}

Since at three-loop order the exact result is not known a criterion is needed
in order to select the best approximation among the various choices at
hand. For this reason we define
\begin{eqnarray}
  \delta_{\rm app} &=& 
  \left|
    \frac{\zeta^{(2)}_{\rm app}-\zeta^{(2)}_{\rm exact}}{\zeta^{(2)}_{\rm exact}}
  \right|
  +
  \left|
    \frac{\zeta^{(3)c}_{\rm app}-\zeta^{(3)}_{\rm app}}{\zeta^{(3)}_{\rm app}}
  \right|\,,
  \label{eq::delta_app}
\end{eqnarray}
where ``app'' marks an approximation result and the superscript ``c''
indicates that the highest terms in the expansions are cut.  For each set of
input parameters we choose the representation which leads to the minimal value
of $\delta_{\rm app}$.  The first term on the right-hand side of
Eq.~(\ref{eq::delta_app}) guarantees that the approximation works well at
two-loop order whereas the second term assures the convergence of the
expansion.
  
The three-loop result $\zeta_{\alpha_s}^{(3)}$ is shown in
Fig.~\ref{fig::zeta3l} as a function of $M_{\tilde{u}_3,R}$.  The notation for
the three hierarchies is as in Fig.~\ref{fig::zeta2l}.  The thick lines are
obtained using all available expansion terms whereas for the thin curves the
highest order is set to zero. Thus the difference between the thick and the
corresponding thin lines is a measure for the quality of the convergence.

One observes a smiliar behaviour as at two-loop order: For small values of
$M_{\tilde{u}_3,R}$, which correspond to small values of $\mstopone$, both
(h2) and (h3) provide good approximations. With increasing $M_{\tilde{u}_3,R}$
(h3) becomes worse whereas $\zeta_{\rm app}^{(3)}$ and $\zeta_{\rm
  app}^{(3)c}$ for (h2) are still practically on top of each other. For values
300~GeV$\lsim M_{\tilde{u}_3,R} \lsim$800~GeV the top squark masses are
relatively close to each other which is the region of validity for (h1).  For
higher values one observes again a strong hierarchy between $\mstopone$ and
$\mstoptwo$ and thus (h2) takes over.  It is interesting to note that for each
value of $M_{\tilde{u}_3,R}$ there is at least one hierarchy with a small
value of $\delta_{\rm app}$ and thus an expected good approximation to the
unknown exact result. Furthermore, the approximations show a significant
overlap so that the whole range of $M_{\tilde{u}_3,R}$ is covered.

\begin{figure}[t]
  \centering
  \includegraphics[width=1.\linewidth]{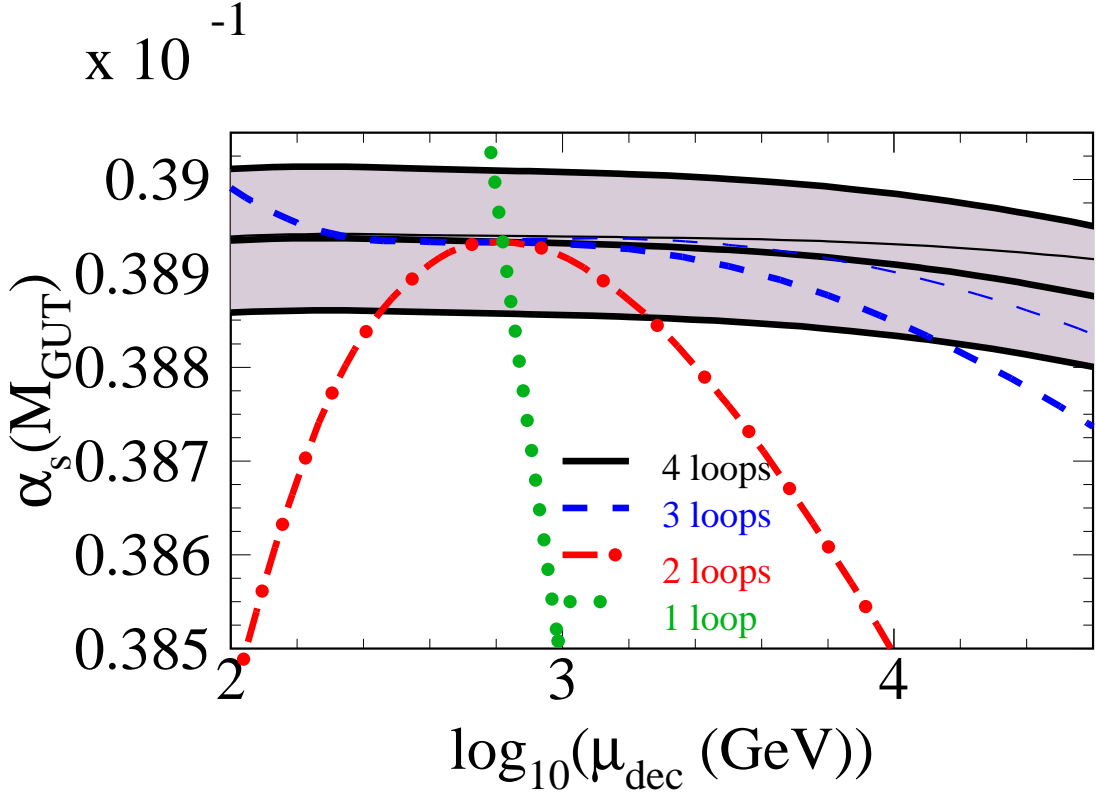}
  \caption[]{\label{fig::asGUT}$\alpha_s^{\rm (SQCD)}(M_{\rm GUT})$ as a function
    of $\mu_{\rm dec}$. Thick and thin 
    lines correspond to the one- and two-step scenario, respectively.
  Thin lines are only shown for three- and four-loop running.}
\end{figure}

Let us in the following briefly discuss the numerical impact of the three-loop
corrections computed in this paper.  In Figs.~\ref{fig::asGUT} we show the
strong coupling at the GUT scale, $\alpha_s^{\rm (SQCD)}(M_{\rm GUT})$ with 
$M_{\rm GUT}=2\cdot10^{16}$~GeV as a
function of the decoupling scale $\mu_{\rm dec}$ which is obtained by the
following procedure.  The starting point is $\alpha_s^{(5),\overline{\rm
    MS}}(M_Z)$.  In a first step we run in the SM from $\mu=M_Z$ to
$\mu=\mu_{\rm dec}$ where the decoupling of the top quark and the SUSY
particles is performed simultaneously and $\alpha_s^{(5)}(\mu_{\rm dec})$ is
transformed to $\alpha_s^{(\rm SQCD)}(\mu_{\rm dec})$.  The use of the SQCD
$\beta$ function finally leads to $\alpha_s^{\rm (SQCD)}(M_{\rm GUT})$.  The
thick lines in Fig.~\ref{fig::asGUT} correspond to this procedure, i.e., we
use the following chain in order to arrive at $\alpha_s^{\rm (SQCD)}(M_{\rm
  GUT})$
\begin{eqnarray}
  \alpha_s^{(5),\overline{\rm MS}}(M_Z)
  \stackrel{\mbox{run.}}{\to} \alpha_s^{(5),\overline{\rm MS}}(\mu_{\rm dec})
  \stackrel{\mbox{dec.}}{\to}  \alpha_s^{(\rm SQCD)}(\mu_{\rm dec})
  \stackrel{\mbox{run.}}{\to}  \alpha_s^{(\rm SQCD)}(M_{\rm GUT})
  \,.
  \label{eq::chain1}
\end{eqnarray}
For a degenerate supersymmetric mass spectrum the decoupling constant can
be found in Eq.~(\ref{eq::zeta_deg}).

Alternatively, in order to obtain the thin lines we integrate out the top
quark in a separate step with $\mu=M_t$ ($M_t$ is the on-shell top quark
mass) and transform afterwards $\alpha_s^{(6),\overline{\rm MS}}$ to
$\alpha_s^{(\rm SQCD)}(M_{\rm GUT})$ in analogy to
Eq.~(\ref{eq::chain1}). Thus we have
\begin{eqnarray}
  &&\alpha_s^{(5),\overline{\rm MS}}(M_Z)
  \stackrel{\mbox{run.}}{\to} \alpha_s^{(5),\overline{\rm MS}}(M_t)
  \stackrel{\mbox{dec.}}{\to} \alpha_s^{(6),\overline{\rm MS}}(M_t)
  \nonumber\\&&
  \stackrel{\mbox{run.}}{\to} \alpha_s^{(6),\overline{\rm MS}}(\mu_{\rm dec})
  \stackrel{\mbox{dec.}}{\to} \alpha_s^{(\rm SQCD)}(\mu_{\rm dec})
  \stackrel{\mbox{run.}}{\to} \alpha_s^{(\rm SQCD)}(M_{\rm GUT})
  \,.
  \label{eq::chain2}
\end{eqnarray}
The decoupling constant needed for the transition from
$\alpha_s^{(6),\overline{\rm MS}}$ to $\alpha_s^{(\rm SQCD)}$
in the limit of degenerate SUSY masses is given in Eq.~(\ref{eq::zeta_deg2}).

In order to obtain the numerical results in Fig.~\ref{fig::asGUT} we have 
used the measured result for $\alpha_s^{(5)}(M_Z)$ which
reads~\cite{PDG} 
\begin{eqnarray}
  \alpha_s^{(5)}(M_Z) &=& 0.1184 \pm 0.0007
  \,.
\end{eqnarray}
Furthermore we have adopted a {\tt mSUGRA} scenario with
\begin{align}
  &&m_0 = 700~\mbox{GeV}\,,
  &&m_{1/2} = 600~\mbox{GeV}\,,
  &&\tan\beta = 10\,,
  &&A_0 = 0\,,
  &&\mu_{\rm SUSY} > 0
  \,.
\end{align}
as input for {\tt softsusy}~\cite{Allanach:2001kg} in order to compute the
supersymmetric mass spectrum. Note that there is only a weak dependence
of the general features of our numerical result on the particular spectrum.
However, it is convenient to make use of a spectrum generator in order to
obtain directly the $\overline{\rm DR}$ values for the masses at the scale
$\mu_{\rm dec}$.  To our knowledge the running of the $\overline{\rm DR}$
parameters is only implemented to two-loop accuracy which poses a slight
inconsistency in our analysis. However, this is only an minor effect and does
not influence the main conclusions.  In order to get an impression about the
numerical values for the physical masses we show the $\overline{\rm DR}$
results for a typical scale $\mu_{\rm dec}=1000$~GeV
\begin{align}
  &&m_t = 146.7~\mbox{GeV}\,,
  &&\mstopone = 1022~\mbox{GeV}\,,
  &&\mstoptwo = 1271~\mbox{GeV}\,,
  \nonumber \\
  &&\msquark = 1348~\mbox{GeV}\,,
  &&\mgluino = 1326~\mbox{GeV}\,,
  &&\theta_{t} = 1.26\,.
\end{align}
At three-loop order the best appoximation is provided by the hierarchy
(h1). In fact the quantity $\delta_{\rm app}$ in Eq.~(\ref{eq::delta_app})
takes the value $\delta_{\rm app} = 0.002$.

For consistency $N$-loop running has to be accompanied with $N-1$-loop
decoupling relations. Thus, we can show curves for $N=1,2,3$ and $4$ which
corresponds to the (thick) dotted, dash-dotted, dashed and solid line,
respectively. Within QCD the beta function is known to four-loop
accuracy~\cite{Vermaseren:1997fq,Czakon:2004bu}, however, the supersymmetric
analogue only to three
loops~\cite{Jack:1996vg,Pickering:2001aq,Harlander:2009mn}.\footnote{The
  four-loop SQCD $\beta$ function is not yet complete~\cite{Jack:1996cn}.}
As a consequence for the four-loop curve in Fig.~\ref{fig::asGUT} we only
use three-loop running above $\mu_{\rm dec}$.

$\mu_{\rm dec}$ is an
unphysical scale not predicted by theory. Thus, on general grounds, the
dependence on $\mu_{\rm dec}$ has to diminish if higher order corrections are
included. This is clearly visible in Fig.~\ref{fig::asGUT} where the dotted,
dash-dotted, dashed and solid lines correspond to one-, two-, three- and
four-loop running, respectively.  Around the central scale of approximately
1000~GeV all loop orders lead to predictions which are quite close. However, a
variation of $\mu_{\rm dec}$ leads to a relatively strong variation of the
two-loop result which gets stabilized at three-loops and which furthermore
gets to a large extend $\mu_{\rm dec}$ independent at four loops.  Actually,
varying $\mu_{\rm dec}$ between 100~GeV and 10\,000~GeV changes $\alpha_s^{\rm
  (SQCD)}(M_{\rm GUT})$ by only 0.07\%.

It is interesting to compare the variation of the individual curves with
respect to
$\mu_{\rm dec}$ with the experimental uncertainty induced from
$\alpha_s^{(5),\overline{\rm MS}}(M_Z)$ which is indicated by the band around
the four-loop curve. The two-loop prediction is inside the band for 
$300$~GeV~$\lsim\mu_{\rm dec}\lsim 1800$~GeV whereas the three-loop curve 
leaves the band only for $\mu_{\rm dec}\gsim 13\,000$~GeV.
It is also interesting to mention that all higher order corrections are very
small for $\mu_{\rm dec}\approx 650$~GeV.

Note that often $\mu_{\rm dec}=M_Z$ is chosen for the
matching between the SM and the MSSM. This choice leads to strong deviation at
two-loops. At three-loop order the results are already quite stable which is
further supported at four loops.

Let us finally remark on the step-by-step decoupling of the top quark and the
supersymmetric particles. The corresponding three- and four-loop 
results are shown as thin lines in 
Fig.~\ref{fig::asGUT}. One observes even flatter curves than for the one-step
scenario, however, the difference is numerically small and well within
the uncertainty band. In this context we want to stress the wide range of
$\mu_{\rm dec}$ which is considered in Fig.~\ref{fig::asGUT}.


\section{\label{sec::let}Low-energy theorem and Higgs-gluon coupling in
  supersymmetric QCD}

In Ref.~\cite{Chetyrkin:1997un} the following formula valid for all orders in
perturbation theory has been derived
in the framework of QCD\footnote{Note the different normalization of the
  operator ${\cal O}_1$ in Ref.~\cite{Chetyrkin:1997un}.}
\begin{eqnarray}
  C_1 &=& D_h^{\rm QCD} \ln\zeta_{\alpha_s}
  \label{eq::LETQCD}
\end{eqnarray}
where 
\begin{eqnarray}
  D_h^{\rm QCD} &=& - m_h \frac{\partial}{\partial m_h}
  \label{eq::QCD}
\end{eqnarray}
describes the derivative with respect to the heavy mass $m_h$.  Thus the $N$-loop
corrections to $\zeta_{\alpha_s}$ immediately leads to $N$-loop corrections to
$C_1$. Since in Eq.~(\ref{eq::LETQCD}) a logarithmic derivative is taken and
furthermore the dependence of $\zeta_{\alpha_s}$ on $m_h$ only occurs via
$\ln(\mu^2/m_h^2)$ even the $(N+1)$ corrections of $C_1$ can be computed once
the renormalization scale dependence of $\zeta_{\alpha_s}$ at $(N+1)$-loop
order is re-constructed with the help of the renormalization group equations.

The LET of Eq.~(\ref{eq::LETQCD}) can easily be extended to the case where
more than one heavy quark is present. This version has been used in
Ref.~\cite{Grozin:2011nk} in order to derive $C_1$ for theories with several
heavy quarks which couple in a Yukawa-like way to the Higgs boson.

The extension of Eq.~(\ref{eq::LETQCD}) to NLO corrections in the framework of
the MSSM has been considered in Ref.~\cite{Degrassi:2008zj}.  Because of the
different setup of our calculation, which is mainly due to the $\varepsilon$
scalars, we cannot take over the derivation of Ref.~\cite{Degrassi:2008zj}. However,
following the same line of reasoning as in Ref.~\cite{Chetyrkin:1997un} we obtain a
version of the LET which is appropriate for the decoupling constants computed
in the previous chapter. For this purpose it is convenient to consider
the bare decoupling constant (see Eq.~(\ref{eq::zeta_bare})) expressed
in terms of bare parameters. This leads to the LET in the form
\begin{eqnarray}
  C_1^0 &=& D_h^0 \ln \zeta_{\alpha_s}^0
  \,.
  \label{eq::LET1}
\end{eqnarray}
$D_h^0$ contains derivatives with respect to bare parameters (indicated by the
superscript ``0'') and can be written as
\begin{eqnarray}
  D_h^0 &=& D_{\tilde{t}}^0 + D_{\tilde{q}}^0 
  + V_t^0 \frac{\partial}{\partial m_t^0}
  + \left(\Lambda_\varepsilon^0\right)^2
  \frac{\partial}{\partial \left(m_\varepsilon^0\right)^2}
  \,.
  \label{eq::DhSUSY}
\end{eqnarray}
$\Lambda_\varepsilon$ is the evanescent Higgs boson-$\varepsilon$ scalar
coupling which is best defined through the corresponding part of the Lagrange
density~\cite{Pak:2010cu}
\begin{eqnarray}
  {\cal L}_\varepsilon &=& 
  -\frac{1}{2}\left(M_\varepsilon^0\right)^2
  \varepsilon^{0,a}_\sigma\varepsilon^{0,a}_\sigma 
  - \frac{\phi^0}{v^0} \left(\Lambda_\varepsilon^0\right)^2
  \varepsilon^{0,a}_\sigma\varepsilon^{0,a}_\sigma 
  \,.
  \label{eq::Lep}
\end{eqnarray}
For convenience we have also displayed the mass term for the $\varepsilon$ scalar.

The derivative operators in Eq.~(\ref{eq::DhSUSY}) are defined through\footnote{In
  order to keep the notation simple we omit the superscript ``0'' in these
  expressions.}
\begin{align*}
  & D_{\tilde{t}} = V_{11}^{\tilde{t}} \frac{\partial}{\partial
    m^2_{\tilde{t}_1}} + V_{22}^{\tilde{t}} \frac{\partial}{\partial
    m^2_{\tilde{t}_2}} + \frac{V_{12}^{\tilde{t}} +
    V_{21}^{\tilde{t}}}{2(m^2_{\tilde{t}_1} - m^2_{\tilde{t}_2})}
  \frac{\partial}{\partial \theta_t}\,,\\
  & D_{\tilde{q}} = V_{11}^{\tilde{q}} \frac{\partial}{\partial
    m^2_{\tilde{q}_1}} + V_{22}^{\tilde{q}} \frac{\partial}{\partial
    m^2_{\tilde{q}_2}} + \frac{V_{12}^{\tilde{q}} +
    V_{21}^{\tilde{q}}}{2(m^2_{\tilde{q}_1} - m^2_{\tilde{q}_2})}
  \frac{\partial}{\partial \theta_q}\,,
\end{align*}
where the prefactors in the top quark sector are obtained from the relations
\begin{align*}
  & V_t = - m_t \frac{\cos\alpha}{\sin\beta}\,,\\
  & V_{\text{\tiny LL}}^{\tilde{t}} = - 2 m_t^2 \frac{\cos\alpha}{\sin\beta} + M_Z^2 \cos^2\theta_W \left( 1 - \frac{1}{3} \tan^2\theta_W \right) \sin (\alpha + \beta)\,,\\
  & V_{\text{\tiny RR}}^{\tilde{t}} = - 2 m_t^2 \frac{\cos\alpha}{\sin\beta} + \frac{4}{3} M_Z^2 \sin^2\theta_W \sin (\alpha + \beta)\,,\\
  & V_{\text{\tiny LR}}^{\tilde{t}} = V_{\text{\tiny RL}}^{\tilde{t}} = \frac{m_t}{\sin\beta} ( - \mu_{\text{\tiny SUSY}} \sin\alpha - A_t \cos\alpha )\,,\\
  & \begin{pmatrix}V_{11}^{\tilde{t}} & V_{12}^{\tilde{t}} \\
    V_{21}^{\tilde{t}} & V_{22}^{\tilde{t}}\end{pmatrix} =
  R(\theta_t)^\dag \begin{pmatrix}V_{\text{\tiny LL}}^{\tilde{t}} &
    V_{\text{\tiny LR}}^{\tilde{t}} \\ V_{\text{\tiny RL}}^{\tilde{t}} &
    V_{\text{\tiny RR}}^{\tilde{t}}\end{pmatrix} R(\theta_t)\,,
\end{align*}
with
\begin{align*} 
& R(\theta_t) = \begin{pmatrix}\cos\theta_t & -\sin\theta_t \\ \sin\theta_t &
  \cos\theta_t\end{pmatrix}\,.
\end{align*}
All light quark masses are set to zero. Their averaged contribution is denoted
by $q$ and thus we have
\begin{align*}
  & V_{\text{\tiny LL}}^{\tilde{q}} = \frac{1}{n_l} \left( \frac{n_l + n_t}{2} V_{\text{\tiny LL}}^{\tilde{d}} + \frac{n_l - n_t}{2} V_{\text{\tiny LL}}^{\tilde{u}} \right)\,,\\
  & V_{\text{\tiny RR}}^{\tilde{q}} = \frac{1}{n_l} \left( \frac{n_l + n_t}{2} V_{\text{\tiny RR}}^{\tilde{d}} + \frac{n_l - n_t}{2} V_{\text{\tiny RR}}^{\tilde{u}} \right)\,,\\
  & V_{\text{\tiny LR}}^{\tilde{q}} = V_{\text{\tiny RL}}^{\tilde{q}} = 0\,,\\
  & V_{\text{\tiny LL}}^{\tilde{u}} = M_Z^2 \cos^2\theta_W \left( 1 - \frac{1}{3} \tan^2\theta_W \right) \sin (\alpha + \beta)\,,\\
  & V_{\text{\tiny RR}}^{\tilde{u}} = \frac{4}{3} M_Z^2 \sin^2\theta_W \sin (\alpha + \beta)\,,\\
  & V_{\text{\tiny LL}}^{\tilde{d}} = M_Z^2 \cos^2\theta_W \left( - 1 - \frac{1}{3} \tan^2\theta_W \right) \sin (\alpha + \beta)\,,\\
  & V_{\text{\tiny RR}}^{\tilde{d}} = - \frac{2}{3} M_Z^2 \sin^2\theta_W \sin
  (\alpha + \beta)\,,
\end{align*}
where ``$u$'' and ``$d$'' denote generic up- and down-type squarks,
respectively, and the labels $n_l = 5$ and $n_t = 1$ are kept arbitrary for
convenience.  $V_{ij}^{\tilde{q}}$ with $i,j=1,2$ are obtained in analogy to
$V_{ij}^{\tilde{t}}$.

After applying $D_h^0$ to $\zeta_{\alpha_s}^0$ of Section~\ref{sec::dec} we
obtain the coefficient function $C_1$ expressed in terms of bare parameters.
Thus, in a next step one has to perform the parameter
renormalization.\footnote{The details are described in
  Ref.~\cite{Pak:2010cu,PSZ_2012}.}  Furthermore, it is necessary to take
into account the operator renormalization constant, often denoted by
$Z_{11}$~\cite{Pak:2010cu}, to obtain a finite result for the coefficient
function which can then be compared to~\cite{Pak:2010cu,PSZ_2012}.

An alternative version of the LET (compared to Eq.~(\ref{eq::LET1})) is
obtained by exploiting the fact that $Z_{\alpha_s}$ and
$Z_{\alpha_s^\prime}$ are independent of the parameters occurring in
$D_h^0$. Thus we can write
\begin{eqnarray}
  C_1 &=& D_h^0 \ln \zeta_{\alpha_s}
  \,,
  \label{eq::LET2}
\end{eqnarray}
where it is still understood that $C_1$ and $\zeta_{\alpha_s}$ are expressed
in terms of unrenormalized parameters.
After computing $C_1$ with the help of Eq.~(\ref{eq::LET2}) the
parameters have to be renormalized as before, however, the operator
renormalization constant is not necessary anymore.

A third version of the LET reads
\begin{eqnarray}
  C_1 &=& D_h \ln \zeta_{\alpha_s}
  \,.
  \label{eq::LET3}
\end{eqnarray}
In this equation all quantities are expressed in terms of $\overline{\rm DR}$
renormalized quantities and $\alpha_s^{\rm SQCD}$, except the evanescent
couplings ($M_{\varepsilon}$ and $\Lambda_\varepsilon$) which are
renormalized to zero. It is very convenient to use Eq.~(\ref{eq::LET3})
since it directly leads to a finite result for $C_1$. It is worth noting that
the computation of $C_1$ from Eq.~(\ref{eq::LET3}) avoids the introduction of
the evanescent coupling $\Lambda_\varepsilon$. This can be understood by
considering the renormalized version of $D_h$ in Eq.~(\ref{eq::DhSUSY}) where
the last term vanishes due to the condition $\Lambda_\varepsilon^2 =
\left(\Lambda_\varepsilon^0\right)^2 - \delta \Lambda_\varepsilon^2 =0$.

Due to the derivatives in Eq.~(\ref{eq::DhSUSY}) the
expansion depth available for $\zeta_{\alpha_s}$ 
is reduced. Nevertheless we can compare the results to
the findings of Ref.~\cite{Pak:2010cu,PSZ_2012} where $C_1$ has been computed
from vertex diagrams. For all three hierarchies we found complete agreement
for the first three terms in the mass difference, i.e. up to order
$(m_i^2-m_j^2)^2$. Furthermore, for (h1) [(h3)]
terms up to $1/\msusy^{6}$ [$1/\msusy^{4}$]  could be compared
successfully and for (h2) all terms including ${\cal O}(1/\mstopone^{4})$ and
${\cal O}(1/\msusy^{4})$ agree.  Thus the calculation of the decoupling
constant together with the application of the LET provides an independent
confirmation of the Higgs-gluon coupling at three-loop order.

The LET in Eq.~(\ref{eq::LET1}) differs from the one presented
in~\cite{Degrassi:2008zj} by the term involving $\Lambda_\varepsilon^0$ (see
Eq.~(\ref{eq::DhSUSY})). Up to NLO it is possible to avoid such a
contribution~\cite{Harlander:2004tp,Degrassi:2008zj}, at three-loop order,
however, a renormalization of the Higgs boson-$\varepsilon$ scalar coupling is
mandatory (see Ref.~\cite{Pak:2010cu,PSZ_2012} for a detailed discussion) in
case derivatives with respect to bare parameters are taken.


\section{\label{sec::concl}Conclusions}

In this paper we have computed the three-loop SQCD corrections to the
decoupling constant relating $\alpha_s$ defined in full MSSM to the one
defined in QCD. The occurring three-loop integrals have been evaluated by
applying expansions in various hierarchies and thus results are obtained which
are valid in a large part of the parameter space.  The decoupling constant
constitutes an important ingredient in the relation of $\alpha_s(M_Z)$ and
$\alpha_s(M_{\rm GUT})$. We have shown that the inclusion three-loop terms to
the decoupling constant in combination with four-loop corrections to the
$\beta$ function leads to results for $\alpha_s(M_{\rm GUT})$ which are
practically independent of the decoupling scale $\mu_{\rm dec}$, where the
effective theory is matched to the full one, even when considering a variation
of $\mu_{\rm dec}$ by more than two orders of magnitude.

A further interesting application of the decoupling constant is its relation
to the effective Higgs-gluon coupling $C_1$ which is obtained by simple
derivatives with respect to the involved parameters. This calculation
constitutes an independent check of the results obtaines in
Ref.~\cite{Pak:2010cu,PSZ_2012} by an explicit calculation. In this paper we
provide the corresponding LET which contains all features also present at
higher orders in perturbation theory. 
It is valid to all orders in perturbation theory.
We have checked that the
renormalized version (cf. Eq.~(\ref{eq::LET3})) works including three-loop
SQCD corrections.



\section*{Acknowledgements}

This work was supported by the DFG through the SFB/TR~9 ``Computational
Particle Physics'' and by the European Commission through the contract
PITN-GA-2010-264564 (LHCPhenoNet).
We would like to thank Konstantin Chetyrkin and Luminita Mihaila for useful
discussions and Luminita Mihaila for carefully reading the manuscript.


\begin{appendix}


\section*{Appendix: Exact one- and two-loop result for $\zeta_{\alpha_s}$}

In this Section we present the results for 
$\zeta_{\alpha_s}$ up to two loops taking into
account the exact dependence on the occurring masses. All parameters are
renormalized in the $\overline{\rm DR}$ scheme except $M_\varepsilon$ which is
renormalized on-shell. 

In contrast to Eq.~(\ref{eq::zetaasexp}) the coefficients of
$\alpha_s^{(5)}$ defined through
\begin{eqnarray}
  \zeta_{\alpha_s}(\mu) &=& 1 
  + \frac{\alpha_s^{(5)}}{\pi} \tilde\zeta_{\alpha_s}^{(1)}
  + \left(\frac{\alpha_s^{(5)}}{\pi}\right)^2 \tilde\zeta_{\alpha_s}^{(2)}
  + \ldots\,,
\end{eqnarray}
is presented. The results read
\begin{align}
  \tilde\zeta_{\alpha_s}^{(1)} &=
  -\frac{1}{4} \Bigg\{C_A \Bigg[ 
\frac{1}{3} 
+\frac{2}{3}l_{\tilde{g}}
\Bigg] 
+  T_F \Bigg[ 
  N_t \Bigg( 
            \frac{1}{3}l_{\tilde{t}_1}
            +\frac{1}{3}l_{\tilde{t}_2} 
            +\frac{4}{3}l_{t}\Bigg) 
            +\frac{2 N_q }{3}l_{\tilde{q}}
      \Bigg] \nonumber\\& 
+\epsilon \Bigg[ T_F \Bigg( 
          N_t \bigg\{
           \frac{1}{6} l_{\tilde{t}_1}^2
          +\frac{1}{6} l_{\tilde{t}_2}^2
          +\frac{2}{3}l_{t}^2 + \zeta_2
           \bigg\} 
        + N_q \bigg\{
        +\frac{1}{3} l_{\tilde{q}}^2
        +\frac{1}{3}\zeta_2\bigg\} \Bigg) \nonumber\\&
 +C_A \Bigg( 
      \frac{1}{3} L_{\epsilon}
     +\frac{1}{3} l_{\tilde{g}}^2
     +\frac{1}{3}\zeta_2
       \Bigg) 
      \Bigg] \Bigg\}\,,\nonumber\displaybreak[1]\\&
  \nonumber\\
  \tilde\zeta_{\alpha_s}^{(2)} &=
 \frac{1}{16}\Bigg\{
 C_A^2 \Bigg[ 
   -\frac{7}{36} 
  - \frac{2}{3}l_{\tilde{g}}
    \Bigg]
+ C_A T_F \Bigg[ N_q \Bigg( 
         \frac{5}{9} 
        + \frac{2 m_{\tilde{g}}^2 }{3 \mathcal{D}_{\tilde{q}\tilde{g}}} l_{\tilde{g}}
        - \frac{2 m_{\tilde{g}}^2 }{3 \mathcal{D}_{\tilde{q}\tilde{g}}}l_{\tilde{q}}
         \Bigg) \nonumber\displaybreak[1]\\&
        + N_t \Bigg( 
        1 
        + \frac{4 \mathcal{N}_{11\,\tilde{t}_1}}{3 \mathcal{D}_{\tilde{t}_1}} 
        +\frac{4 m_{\tilde{g}}^2 m_{\tilde{t}_1}^2 m_t^2 \mathcal{N}_{5\,\tilde{t}_1}}{\mathcal{D}_{\tilde{t}_1}^3}\Phi(m_t,m_{\tilde{t}_1},m_{\tilde{g}})\nonumber\displaybreak[1]\\& 
        - \frac{2 \mathcal{N}_{3\,\tilde{t}_1} }{3 \mathcal{D}_{\tilde{t}_1}^2}l_{\tilde{t}_1} 
        + \bigg[ 
                -\frac{8}{3} 
                + \frac{16 m_{\tilde{g}}^2 m_{\tilde{t}_1}^2 \mathcal{N}_{15\,\tilde{t}_1}}{\mathcal{D}_{\tilde{t}_1}^2}  
                - \frac{2 \mathcal{N}_{21\,\tilde{t}_1}}{3 \mathcal{D}_{\tilde{t}_1}}
          \bigg] l_{t}
        + \bigg[ 
                \frac{2 m_{\tilde{g}}^2 \mathcal{N}_{19\,\tilde{t}_1}}{3 \mathcal{D}_{\tilde{t}_1}} 
               - \frac{8 m_{\tilde{g}}^2 m_{\tilde{t}_1}^2 \mathcal{N}_{6\,\tilde{t}_1}}{\mathcal{D}_{\tilde{t}_1}^2} 
          \bigg] l_{\tilde{g}} \nonumber\displaybreak[1]\\& 
     + c_{\theta_t}s_{\theta_t}\bigg[ 
                          -\frac{4 m_{\tilde{g}} m_t \mathcal{N}_{1\,\tilde{t}} \mathcal{N}_{2\,\tilde{t}}}{3 \mathcal{D}_{\tilde{t}_1} \mathcal{D}_{\tilde{t}_2}} 
                         - \frac{8 m_{\tilde{g}} m_{\tilde{t}_1}^2 m_t \mathcal{N}_{1\,\tilde{t}_1} }{\mathcal{D}_{\tilde{t}_1}^3}\Phi(m_t,m_{\tilde{t}_1},m_{\tilde{g}})   
                         + \frac{8 m_{\tilde{t}_1}^2 m_t \mathcal{N}_{7\,\tilde{t}_1} }{3 \mathcal{D}_{\tilde{t}_1}^2 m_{\tilde{g}}}l_{\tilde{t}_1}\nonumber\displaybreak[1]\\& 
                         + \bigg(
                                   \frac{8 m_t \mathcal{N}_{8\,\tilde{t}_1}}{3 \mathcal{D}_{\tilde{t}_1} m_{\tilde{g}}}
                                   -\frac{16 m_{\tilde{g}} m_{\tilde{t}_1}^2 m_t \mathcal{N}_{6\,\tilde{t}_1}}{\mathcal{D}_{\tilde{t}_1}^2}
                           \bigg) l_{t}
                         - \bigg[
                                  \frac{8 m_{\tilde{g}}^3 m_t}{3 \mathcal{D}_{\tilde{t}_1}} 
                                + \frac{16 m_{\tilde{g}} m_{\tilde{t}_1}^2 m_t}{\mathcal{D}_{\tilde{t}_1}} 
                                + \frac{32 m_{\tilde{g}}^3 m_{\tilde{t}_1}^2 m_t^3}{\mathcal{D}_{\tilde{t}_1}^2}
                           \bigg] l_{\tilde{g}}
                    \bigg] 
       \Bigg) 
      \Bigg] \nonumber\displaybreak[1]\\
& +C_F T_F \Bigg[ N_q \Bigg( \frac{13}{6} 
 - \frac{2 M_{\epsilon}^2}{3 m_{\tilde{q}}^2} 
+ \frac{4 m_{\tilde{g}}^2}{3 m_{\tilde{q}}^2} 
 + \frac{4 m_{\tilde{g}}^4 l_{\tilde{g}}}{3 m_{\tilde{q}}^2 \mathcal{D}_{\tilde{q}\tilde{g}}}
 - 2 l_{\tilde{q}} 
 - \frac{4 m_{\tilde{g}}^2 l_{\tilde{q}}}{3 \mathcal{D}_{\tilde{q}\tilde{g}}}\Bigg)\nonumber\displaybreak[1]\\& 
 + N_t \Bigg( 
              \frac{2}{3}
              - \frac{2 M_{\epsilon}^2}{3 m_{\tilde{t}_1}^2}
              + \frac{4 m_{\tilde{g}}^2}{3 m_{\tilde{t}_1}^2} 
              + \frac{4 m_t^2}{3 m_{\tilde{t}_1}^2}
              - \frac{4 m_{\tilde{g}}^2 \mathcal{N}_{13\,\tilde{t}_1}}{3 \mathcal{D}_{\tilde{t}_1}}
              - \frac{8 m_{\tilde{g}}^4 m_{\tilde{t}_1}^2 m_t^2 \mathcal{N}_{13\,\tilde{t}_1}}{\mathcal{D}_{\tilde{t}_1}^3}\Phi(m_t,m_{\tilde{t}_1},m_{\tilde{g}})\nonumber\displaybreak[1]\\& 
              + \bigg[ 
                          \frac{4 m_{\tilde{g}}^2 m_{\tilde{t}_1}^2}{3 \mathcal{D}_{\tilde{t}_1}} 
                        + \frac{16 m_{\tilde{g}}^4 m_{\tilde{t}_1}^2 m_t^2}{\mathcal{D}_{\tilde{t}_1}^2}
                        -\frac{5}{3} 
                \bigg] l_{\tilde{t}_1} \nonumber\\&  
              - \frac{5}{3} l_{\tilde{t}_2}
               + \bigg[ 2 
              + \frac{4 m_{\tilde{g}}^2}{3 m_{\tilde{t}_1}^2} 
              + \frac{4 m_t^2}{3 m_{\tilde{t}_1}^2}
               + \frac{8 m_{\tilde{g}}^4 \mathcal{N}_{12\,\tilde{t}_1}}{\mathcal{D}_{\tilde{t}_1}^2}
              - \frac{4 m_{\tilde{g}}^2 \mathcal{N}_{9\,\tilde{t}_1}}{3 \mathcal{D}_{\tilde{t}_1} m_{\tilde{t}_1}^2}\bigg] l_{t}\nonumber\displaybreak[1]\\& 
               + \bigg[  
                        \frac{4 m_{\tilde{g}}^4 \mathcal{N}_{20\,\tilde{t}_1}}{3 \mathcal{D}_{\tilde{t}_1} m_{\tilde{t}_1}^2}
                       -\frac{8 m_{\tilde{g}}^4 \mathcal{N}_{11\,\tilde{t}_1}}{\mathcal{D}_{\tilde{t}_1}^2}  
                 \bigg] l_{\tilde{g}}\nonumber\displaybreak[1]\\&
              + s_{\theta_t}c_{\theta_t} 
               \bigg[ 
                      \frac{16 m_{\tilde{g}}^3 m_t}{3 \mathcal{D}_{\tilde{t}_1}} 
                      - \frac{8 m_{\tilde{g}} m_t}{3 m_{\tilde{t}_1}^2}
                       + \frac{16 m_{\tilde{g}}^3 m_{\tilde{t}_1}^2 m_t \mathcal{N}_{14\,\tilde{t}_1}}{\mathcal{D}_{\tilde{t}_1}^3}\Phi(m_t,m_{\tilde{t}_1},m_{\tilde{g}})\nonumber\displaybreak[1]\\& 
                      - \frac{16 m_{\tilde{g}} m_{\tilde{t}_1}^2 \mathcal{N}_{10\,\tilde{t}_1}}{3 \mathcal{D}_{\tilde{t}_1}^2 m_t}l_{\tilde{t}_1} 
                       + \bigg(
                                \frac{8 m_{\tilde{g}} m_t \mathcal{N}_{2\,\tilde{t}_1}}{3 \mathcal{D}_{\tilde{t}_1} m_{\tilde{t}_1}^2}
                               -\frac{8 m_{\tilde{g}} m_t}{3 m_{\tilde{t}_1}^2} 
                              -\frac{16 m_{\tilde{g}}^3 m_t \mathcal{N}_{11\,\tilde{t}_1}}{\mathcal{D}_{\tilde{t}_1}^2}
                         \bigg) l_{t}\nonumber\displaybreak[1]\\&
                      + \bigg( 
                                  \frac{ 16 m_{\tilde{g}}^5 m_t \mathcal{N}_{16\,\tilde{t}_1}}{\mathcal{D}_{\tilde{t}_1}^2}
                                + \frac{8 m_{\tilde{g}}^3 \mathcal{N}_{1\,\tilde{t}}}{3 m_t \mathcal{N}_{18\,\tilde{t}_1} \mathcal{N}_{18\,\tilde{t}_2}} 
                                - \frac{8 m_{\tilde{g}}^3 \mathcal{N}_{4\,\tilde{t}_1}}{3 \mathcal{D}_{\tilde{t}_1} m_{\tilde{t}_1}^2 \mathcal{N}_{18\,\tilde{t}_1}}
                        \bigg) l_{\tilde{g}}
               \bigg]  \nonumber\displaybreak[1]\\& 
              + \Big(s_{\theta_t}^2-s_{\theta_t}^4\Big)\bigg[ 
                        -\frac{2 \mathcal{N}_{1\,\tilde{t}}^2}{3 m_{\tilde{t}_1}^2 m_{\tilde{t}_2}^2}
                        + \bigg( 
                                 \frac{4}{3} 
                                - \frac{4 m_{\tilde{t}_1}^2}{3 m_{\tilde{t}_2}^2}
                          \bigg) l_{\tilde{t}_1}
                \bigg]  
   \Bigg) \Bigg] \Bigg\}
  +
  \Bigg\{\begin{array}{c}  m_{\tilde{t}_2}\leftrightarrow m_{\tilde{t}_1}\\
    \theta_t\rightarrow-\theta_t \end{array}\Bigg\}
  \label{MSSM:ZETAG:EQ:zetaalphas}
  \,,
\end{align}
where $C_F=4/3$, $C_A=3$, $T_F=1/2$, $N_t=1$, $N_q=5$, $\zeta_n$ is the
Riemann zeta function,
$l_x=\ln(\mu^2/m_x^2)$, $L_{\epsilon}=\ln(\mu^2/M_{\epsilon}^2)$ and
$M_{\epsilon}$ is the $\epsilon$ scalar mass. Furthermore we have
\newcommand{\lp}{\left(}
\newcommand{\rp}{\right)}
\begin{align}
\mathcal{D}_{\tilde{t}_i} =&m_{\tilde{g}}^4 + \lp m_{\tilde{t}_i}^2 - m_t^2\rp ^2 - 2 m_{\tilde{g}}^2 \lp m_{\tilde{t}_i}^2 + m_t^2\rp  \,,\nonumber\\
\mathcal{D}_{\tilde{q}\tilde{g}} =&  m_{\tilde{g}}^2 -  m_{\tilde{q}}^2  \,,\nonumber\\ 
\mathcal{N}_{1\,\tilde{t}} =&m_{\tilde{t}_1}^2 - m_{\tilde{t}_2}^2  \,,\nonumber\\
\mathcal{N}_{2\,\tilde{t}} =&m_{\tilde{g}}^4 + m_t^2 \lp m_{\tilde{t}_2}^2 - 3 m_t^2\rp - m_{\tilde{g}}^2 \lp m_{\tilde{t}_1}^2 + m_{\tilde{t}_2}^2 - 2 m_t^2\rp + m_{\tilde{t}_1}^2 \lp m_{\tilde{t}_2}^2 + m_t^2\rp   \,,\nonumber\\ 
\mathcal{N}_{1\,\tilde{t}_i} =&m_{\tilde{g}}^6 - \lp m_{\tilde{t}_i}^2 - m_t^2\rp ^2 \lp m_{\tilde{t}_i}^2 + m_t^2\rp - m_{\tilde{g}}^4 \lp 3 m_{\tilde{t}_i}^2 + m_t^2\rp + m_{\tilde{g}}^2 \lp 3 m_{\tilde{t}_i}^4 + m_t^4\rp   \,,\nonumber\\ 
\mathcal{N}_{2\,\tilde{t}_i} =&m_{\tilde{g}}^4 - 2 m_{\tilde{t}_i}^4 + 2 m_{\tilde{t}_i}^2 m_t^2 + m_{\tilde{g}}^2 \lp m_{\tilde{t}_i}^2 - m_t^2\rp   \,,\nonumber\\
\mathcal{N}_{3\,\tilde{t}_i} =&m_{\tilde{g}}^8 - m_{\tilde{g}}^6 \lp 3 m_{\tilde{t}_i}^2 + 4 m_t^2\rp + \lp -3 m_{\tilde{t}_i}^2 + m_t^2\rp \lp - m_{\tilde{t}_i}^2 m_t + m_t^3\rp ^2 \nonumber\\
                              &+ m_{\tilde{g}}^4 \lp 3 m_{\tilde{t}_i}^4 + 13 m_{\tilde{t}_i}^2 m_t^2 + 6 m_t^4\rp - m_{\tilde{g}}^2 \lp m_{\tilde{t}_i}^6 + 6 m_{\tilde{t}_i}^4 m_t^2 + 5 m_{\tilde{t}_i}^2 m_t^4 + 4 m_t^6\rp\,,\nonumber\\ 
\mathcal{N}_{4\,\tilde{t}_i} =&\lp m_{\tilde{g}}^2 - 3 m_{\tilde{t}_i}^2\rp m_t \lp m_{\tilde{g}}^2 + 3 m_{\tilde{t}_i}^2 - m_t^2\rp   \,,\nonumber\\ 
\mathcal{N}_{5\,\tilde{t}_i} =&m_{\tilde{g}}^4 - 2 m_{\tilde{g}}^2 m_{\tilde{t}_i}^2 + m_{\tilde{t}_i}^4 - m_t^4  \,,\nonumber\\ 
\mathcal{N}_{6\,\tilde{t}_i} =&m_{\tilde{g}}^4 + m_{\tilde{t}_i}^4 - m_{\tilde{t}_i}^2 m_t^2 - m_{\tilde{g}}^2 \lp 2 m_{\tilde{t}_i}^2 + 3 m_t^2\rp   \,,\nonumber\\ 
\mathcal{N}_{7\,\tilde{t}_i} =&5 m_{\tilde{g}}^6 + 2 \lp m_{\tilde{t}_i}^2 - m_t^2\rp ^3 - 2 m_{\tilde{g}}^4 \lp 4 m_{\tilde{t}_i}^2 + 3 m_t^2\rp + m_{\tilde{g}}^2 \lp m_{\tilde{t}_i}^4 - 4 m_{\tilde{t}_i}^2 m_t^2 + 3 m_t^4\rp   \,,\nonumber\\ 
\mathcal{N}_{8\,\tilde{t}_i} =&m_{\tilde{g}}^4 + 7 m_{\tilde{g}}^2 m_{\tilde{t}_i}^2 - 2 m_{\tilde{t}_i}^4 + 2 m_{\tilde{t}_i}^2 m_t^2  \,,\nonumber\\ 
\mathcal{N}_{9\,\tilde{t}_i} =&m_{\tilde{g}}^4 + m_{\tilde{t}_i}^4 + m_{\tilde{g}}^2 \lp 4 m_{\tilde{t}_i}^2 - m_t^2\rp   \,,\nonumber\\ 
\mathcal{N}_{10\,\tilde{t}_i} =&m_{\tilde{g}}^6 - \lp m_{\tilde{t}_i}^2 - m_t^2\rp ^3 + m_{\tilde{g}}^4 \lp -3 m_{\tilde{t}_i}^2 + 2 m_t^2\rp + m_{\tilde{g}}^2 \lp 3 m_{\tilde{t}_i}^4 - 5 m_{\tilde{t}_i}^2 m_t^2 + 2 m_t^4\rp   \,,\nonumber\\ 
\mathcal{N}_{11\,\tilde{t}_i} =&m_{\tilde{g}}^4 + m_{\tilde{t}_i}^4 - m_{\tilde{t}_i}^2 m_t^2 - m_{\tilde{g}}^2 \lp 2 m_{\tilde{t}_i}^2 + m_t^2\rp   \,,\nonumber\\ 
\mathcal{N}_{12\,\tilde{t}_i} =&m_{\tilde{g}}^4 + m_{\tilde{t}_i}^4 - 3 m_{\tilde{t}_i}^2 m_t^2 - m_{\tilde{g}}^2 \lp 2 m_{\tilde{t}_i}^2 + m_t^2\rp   \,,\nonumber\\ 
\mathcal{N}_{13\,\tilde{t}_i} =&m_{\tilde{g}}^2 - m_{\tilde{t}_i}^2 + m_t^2  \,,\nonumber\\ 
\mathcal{N}_{14\,\tilde{t}_i} =&m_{\tilde{g}}^4 - 2 m_{\tilde{g}}^2 m_{\tilde{t}_i}^2 + \lp m_{\tilde{t}_i}^2 - m_t^2\rp ^2  \,,\nonumber\\ 
\mathcal{N}_{15\,\tilde{t}_i} =&m_{\tilde{g}}^4 + m_{\tilde{t}_i}^4 - 2 m_{\tilde{t}_i}^2 m_t^2 - 2 m_{\tilde{g}}^2 \lp m_{\tilde{t}_i}^2 + m_t^2\rp   \,,\nonumber\\ 
\mathcal{N}_{16\,\tilde{t}_i} =&m_{\tilde{g}}^2 - m_{\tilde{t}_i}^2 - m_t^2  \,,\nonumber\\ 
\mathcal{N}_{18\,\tilde{t}_i} =& m_{\tilde{g}}^2 - m_{\tilde{t}_i}^2\,,\nonumber\\ 
\mathcal{N}_{19\,\tilde{t}_i} =&m_{\tilde{g}}^2 + 11 m_{\tilde{t}_i}^2 + m_t^2  \,,\nonumber\\ 
\mathcal{N}_{20\,\tilde{t}_i} =&m_{\tilde{g}}^2 + 4 m_{\tilde{t}_i}^2 - m_t^2  \,,\nonumber\\
\mathcal{N}_{21\,\tilde{t}_i} =&m_{\tilde{g}}^4 + m_{\tilde{t}_i}^2 \lp m_{\tilde{t}_i}^2 + m_t^2\rp + m_{\tilde{g}}^2 \lp 22 m_{\tilde{t}_i}^2 + m_t^2\rp   \,,\nonumber
\end{align}
where following abbreviations have been introduced
\small
\begin{align}
\lambda(x,\,y) =& \sqrt{(1 - x - y)^2 - 4xy}\,,\nonumber\\
\text{Cl}_2(x)=& \text{Im}\Big[\text{Li}_2( e^{ix})\Big]\,,\nonumber\\
\Phi_1(x,\, y) =& \lambda^{-1}(x,\,y) 
  \Big\{
    2 \ln\big[\tfrac{1}{2}(1 + x - y - \lambda(x,\,y))\big]\ln\big[\tfrac{1}{2}(1 - x + y - \lambda(x,\,y))\big]+ \tfrac{1}{3}\pi^2\nonumber\\
    &- \ln{x} \ln{y} -2 \text{Li}_2[\tfrac{1}{2}(1 + x - y - \lambda(x,\,y))] -2 \text{Li}_2[\tfrac{1}{2}(1 - x + y - \lambda(x,\,y))] 
  \Big\}\,,\nonumber\\
\Phi_2(x,\, y) =&  \frac{2}{\sqrt{-\lambda^2(x,\,y)}}\Big\{
     \text{Cl}_2\big(2 \arccos{\frac{-1 + x + y}{2 \sqrt{x y}}}\big)\nonumber\\&  
    +\text{Cl}_2\big(2 \arccos{\frac{1 + x - y}{2 \sqrt{x}}}\big)
    +\text{Cl}_2\big(2 \arccos{\frac{1 - x + y}{2 \sqrt{y}}}\big)\Big\}\,,\nonumber\\
\Phi(m_1,m_2,m_3) =&\begin{cases} 
m_3^2 \lambda^2\Big(\frac{m_1^2}{m_3^2},\,\frac{m_2^2}{m_3^2}\Big)\Phi_2\Big(\frac{m_1^2}{m_3^2},\,\frac{m_2^2}{m_3^2}\Big) & \text{Re}\Big[\lambda^2\Big(\frac{m_1^2}{m_3^2},\, \frac{m_2^2}{m_3^2}\Big)\Big] < 0\\
m_3^2 \lambda^2\Big(\frac{m_1^2}{m_3^2},\,\frac{m_2^2}{m_3^2}\Big)\Phi_1\Big(\frac{m_1^2}{m_3^2},\,\frac{m_2^2}{m_3^2}\Big) &  m_1 + m_2  \leq m_3\\
m_1^2 \lambda^2\Big(\frac{m_2^2}{m_1^2},\,\frac{m_3^2}{m_1^2}\Big)\Phi_1\Big(\frac{m_2^2}{m_1^2},\,\frac{m_3^2}{m_1^2}\Big) &  m_2 + m_3  \leq m_1\\
m_2^2 \lambda^2\Big(\frac{m_1^2}{m_2^2},\,\frac{m_3^2}{m_2^2}\Big)\Phi_1\Big(\frac{m_1^2}{m_2^2},\,\frac{m_3^2}{m_2^2}\Big) &  m_1 + m_3  \leq m_2\,.\nonumber
\end{cases}
\end{align}
The one-loop result agrees with Ref.~\cite{Harlander:2005wm}.  The two-loop
result has also been considered in Ref.~\cite{Bauer:2008bj}, however, no
compact result has been presented. Furthermore, the decoupling has only been
considered within DRED, i.e., the transition from DREG to DRED has been
performed in a separate step.


\end{appendix}



\end{document}